\documentclass{blois}

\pdfoutput=1

\usepackage{amsmath}
\usepackage{amsfonts}
\usepackage{amssymb}

\def\be{\begin{equation}}
\def\ee{\end{equation}}
\def\bea{\begin{eqnarray}}
\def\eea{\end{eqnarray}}

\newcommand{\EB}{E_\mathrm{B}}
\newcommand{\EE}{E_{\theta}}
\newcommand{\vw}{v_{\mathrm{w}}}
\newcommand{\SU}{\mathrm{SU}}
\newcommand{\Sp}{\mathrm{Sp}}
\newcommand{\SUL}{\mathrm{SU}(2)_{\mathrm{L}}}
\newcommand{\SUR}{\mathrm{SU}(2)_{\mathrm{R}}}
\newcommand{\ii}{\mathrm{i}}

\newcommand{\Photo}{}

\begin{document}
\vspace*{4cm}
\title{RADIATIVELY INDUCED FERMI SCALE AND UNIFICATION}

\author{ T. ALANNE }

\address{CP$^{3}$-Origins, University of Southern Denmark, Campusvej 55, DK-5230 Odense M, Denmark}

\maketitle\abstracts{
We propose a framework, where the hierarchy between the unification and the Fermi scale emerges radiatively.
This work tackles the long-standing question about the connection between the low Fermi scale
and a more fundamental scale of Nature. As a concrete example, we study a Pati--Salam-type unification
of Elementary-Goldstone-Higgs scenario, where the Standard Model scalar sector is replaced by an SU(4)-symmetric
one, and the observed Higgs particle is an elementary pseudo-Goldstone boson. We construct a concrete
model where the unification scale is fixed to a phenomenologically viable value, while the Fermi scale is
generated radiatively. This scenario provides an interesting link between the unification and Fermi scale
physics, and opens up prospects for exploring a wide variety of open problems in particle physics, ranging
from neutrinos to cosmic inflation.}

\section{Introduction}

    One of the long-standing paradigms beyond the Standard Model (SM) is the unification of (some of) the SM interactions. The minimal unification of strong and electroweak (EW) interactions was proposed by Georgi and Glashow~\cite{Georgi:1974sy} originally into unified $\SU(5)$ gauge group. Such constructions predict gauge-mediated proton decay, and the current lower bound on the proton lifetime~\cite{Nishino:2012bnw}, $\tau> 10^{34}$~y, sets a lower bound on the unification scale to $\gtrsim 10^{15}$~GeV.  

   An alternative framework was proposed by Pati and Salam~\cite{Pati:1974yy} to unify quarks and leptons by promoting the lepton number to the fourth colour. In this scenario, the proton does not decay via gauge interactions,
   but spin-one leptoquarks mediate $K_{\mathrm{L}}\rightarrow\mu^{\pm} e^{\mp}$. Stringent experimental limits on these decays lead to a lower bound $M>1.5\cdot 10^6$~GeV on the leptoquark masses~\cite{Parida:2014dba} translating into the lower bound on the Pati--Salam unification scale $\Lambda_{\mathrm{PS}}\gtrsim 1.9\cdot 10^6$~GeV.

   In either framework, this implies a large  hierarchy between the Fermi scale, $\vw=246$ GeV, and the unification scale. These symmetry breakings are typically modelled via ad-hoc scalar sectors, and there is no symmetry reason to prohibit the portal interactions between these two sectors. However, unless the portal coupling is highly suppressed compared to the SM Higgs self-coupling, the symmetry breaking at the unification scale would induce a large mass for the SM Higgs already at the tree level.

   Therefore, one common scale for the full scalar sector, with the Fermi scale emerging radiatively, would be an appealing alternative.  We show that such a scenario can arise when the Higgs is an elementary pseudo-Goldstone boson (pGB) related to an enhanced global symmetry~\cite{Alanne:2014kea}.

\section{Vacuum misalignment and emergent Fermi scale}
\label{sec:vacAlign}

    In the presence of enhanced global symmetries, there is a possibility of misalignment between the EW gauge group and the stability group related to the spontaneous global symmetry breaking, and the preferred alignment between these subgroups is determined dynamically by the quantum effects. This was first noticed in the context of technicolour by Peskin~\cite{Peskin:1980gc} and Preskill~\cite{Preskill:1980mz}: They discovered that  EW gauge sector prefers to be unbroken, thereby trying to destabilise the technicolour vacuum. Later, it was noticed that in the technicolour models, the corrections from the SM-fermion sector, most notably the top quark, instead prefer the technicolour vacuum~\cite{Galloway:2010bp,Cacciapaglia:2014uja}. Recently, this discussion was extended to models with elementary scalars~\cite{Alanne:2016mmn}.

In general, the true vacuum is a linear combination of the EW preserving vacuum, $E_0$, and the one that fully breaks the EW symmetry to electromagnetism, $E_{\mathrm{B}}$. It is convenient to parameterise the misalignment of the EW subgroup and the stability group, $H$, by an angle, $\theta$, and write the vacuum as
\begin{equation}
    \label{eq:}
    \EE=\cos\theta E_0+\sin\theta\EB.
\end{equation}
The angle, $\theta$, is a priori a free parameter of the model, but will be determined by radiative corrections. For $\theta=0$, the EW symmetry remains unbroken and for $\theta=\pi/2$ the EW symmetry directly breaks to $\mathrm{U}(1)_Q$.
The amount of the breaking of the EW subgroup then depends on the alignment, and the Fermi scale is given by $\vw=v \sin\theta$.
In particular, if the vacuum aligns to a non-zero but small value of the angle, $\theta\ll 1$, the Fermi scale is much smaller than the actual 
symmetry breaking scale, $v$.

    \subsection{The $\SU(4)\rightarrow\Sp(4)$ breaking pattern}
	The $\SU(4)/\Sp(4)$ model with elementary scalars has been studied in both non-supersymmetric~\cite{Alanne:2014kea,Gertov:2015xma} and supersymmetric frameworks~\cite{Alanne:2016uta}. The breaking pattern is obtained by a scalar transforming in the six-dimensional antisymmetric representation of $\SU(4)$, and can be conveniently parameterised as
	\begin{equation}
	    \label{eq:}
	    M=(\frac{\sigma}{2}+\ii\sqrt{2}\Pi^aX^a)E,
	\end{equation}
	where $X^a$ are the broken generators with respect to vacuum $E$, and $\Pi^a$ the corresponding Goldstone bosons (GBs). The global $\SU(4)$ breaks spontaneously to $\Sp(4)$ when $\sigma$ acquires a vev, $v$, i.e. $\langle M\rangle=\frac{v}{2}E$.

	We embed $\SUL\times\SUR$ into $\SU(4)$ by identifying the generators
	    \begin{equation}
		\label{eq:gensCust}
		T^i_{\mathrm{L}}=\frac{1}{2}\left(\begin{array}{cc}\sigma_i & 0 \\ 0 & 0\end{array}\right),\quad\text{and}\quad
		T^i_{\mathrm{R}}=\frac{1}{2}\left(\begin{array}{cc} 0 & 0 \\ 0 & -\sigma_i^{T}\end{array}\right),
	    \end{equation}
	    where $\sigma_i$ are the Pauli matrices. The generator of the hypercharge is then identified with the third generator 
	    of the $\SU(2)_{\mathrm{R}}$ group, $T_Y=T^3_{\mathrm{R}}$. Then, as discussed above, the misalignment between the EW group and the stability group, $\Sp(4)$, can be parameterised by and angle $\theta$, once we identify the EW preserving and breaking vacua, $E_0$ and $\EB$, resp.:
    \begin{equation}
	E_{0}= \left(\begin{array}{cc} \ii \, \sigma_2 & 0\\
	0 & -\ii \, \sigma_2 \\
	\end{array}\right), \quad \EB=\left(\begin{array}{cc} 
	0 & 1\\
	-1& 0\\
	\end{array}\right).
    \end{equation}
    
 The specific value of $\theta$ is determined dynamically once the EW and top-quark quantum corrections are taken into account. The one-loop potential in the $\overline{\mathrm{MS}}$ scheme is given by
	\begin{equation}
	    \label{eq:deltaV}
	    ~\hspace{-0.35cm}\delta V=\frac{1}{64\pi^2}\mathrm{Str}\left[{\cal M}^4(\Phi)\left(\log\frac{{\cal M}^2(\Phi)}
		{\mu_0^2}-C\right)\right],
	\end{equation}
	where ${\cal M}(\Phi)$ is the tree-level mass matrix in the $\Phi$ background.  The supertrace 
	$\mathrm{Str}$, is defined by $\mathrm{Str} = \sum_{\text{scalars}}-2\sum_{\text{fermions}}+3\sum_{\text{vectors}}$,
	and $\displaystyle{C=3/2}$ for scalars and fermions, while $C=\displaystyle{5/6}$ for the gauge bosons. 
		
These corrections favour small values of the angle~\cite{Alanne:2014kea,Gertov:2015xma}, and consequently the Fermi scale, $v_{w} = v \sin \theta$, lies well below the  spontaneous symmetry breaking scale  $v$. Furthermore, the radiative corrections provide a mass for the pGB Higgs via the Coleman--Weinberg mechanism.

\section{Connecting EWSB and unification in a minimal Pati--Salam framework}

We introduced the minimal framework connecting the vacuum misalignment and unification utilising the global symmetry breaking pattern $\SU(4)\rightarrow\Sp(4)$ in the Pati--Salam framework~\cite{Alanne:2015fqh}. The full symmetry group is then $G=\SU(4)_{\mathrm{LC}}\times\SU(4)_{\chi}$, where the subscript LC refers to leptocolour, and the EW gauge group is embedded as a subgroup of the global $\SU(4)_{\chi}$. As described above, the scalar, $M$, transfroming as $(1,6_{\mathrm{A}})$ under $G$, breaks the EW group, and the minimal extension to incorporate the breaking of the leptocolour group to $\SU(3)_{\mathrm{c}}\times U(1)_{\mathrm{B}-\mathrm{L}}$ amounts to adding another scalar multiplet, $P=p_aT^a$, transforming in the adjoint representation under the leptocolour group, i.e. $P\sim(15,1)$. 

The most general renormalisable scalar potential including these scalar multiplets then reads
    \begin{equation}
	\label{eq:pot}
	\begin{split}
	    V=&\frac{1}{2}m_M^2\mathrm{Tr}[M^{\dagger} M]+m_P^2\mathrm{Tr}[P^{2}]+\frac{\lambda_M}{4}\mathrm{Tr}[M^{\dagger}M]^2
		+\lambda_{P1}\mathrm{Tr}[P^2]^2\\
	    &+\lambda_{P2}\mathrm{Tr}[P^4]+\frac{\lambda_{MP}}{2}\mathrm{Tr}[M^{\dagger}M]\mathrm{Tr}[P^2],
	\end{split}
    \end{equation}
and the desired breaking pattern occurs as the scalars acquire vevs $\langle M \rangle=\frac{v_0}{2}E$, and $\langle P \rangle=b_0 T^{15}$.

On the level of the tree-level scalar potential, the alignment between the EW subgroup and the stability group, $\Sp(4)$, is left undetermined. To find the value of the misalignment angle, we need to minimize the full one-loop effective potential, as described in the Sec.~\ref{sec:vacAlign}. We emphasize that the quantum effects dynamically generate the specific value of $\theta$ at the minimum. In particular, the small value of $\theta$ results from the top-Yukawa interaction and the EW gauge sector that break the global symmetry explicitly. Furthermore, fixing the correct Higgs mass is an important constrain to the parameter space. 

We fix the renormalization scale by requiring that the vev $v=\langle\sigma\rangle$ is given by the tree-level value $v_0$, while the one for $\langle p_{15}\rangle=b$ is determined by minimising the full one-loop potential along with the dynamical value of $\theta$.		
Three states, $\Pi_4$, $ \sigma$ and $p_{15}$, have the same quantum numbers as Higgs, and therefore they mix. 
The constraints from the Higgs phenomenology were investigated previously~\cite{Alanne:2014kea,Gertov:2015xma}, and it was shown that the elementary-Goldstone-Higgs paradigm reproduces the phenomenological success of the SM. The same analysis applies here.

\begin{figure}
	\begin{center}
	    \includegraphics[width=0.44\textwidth]{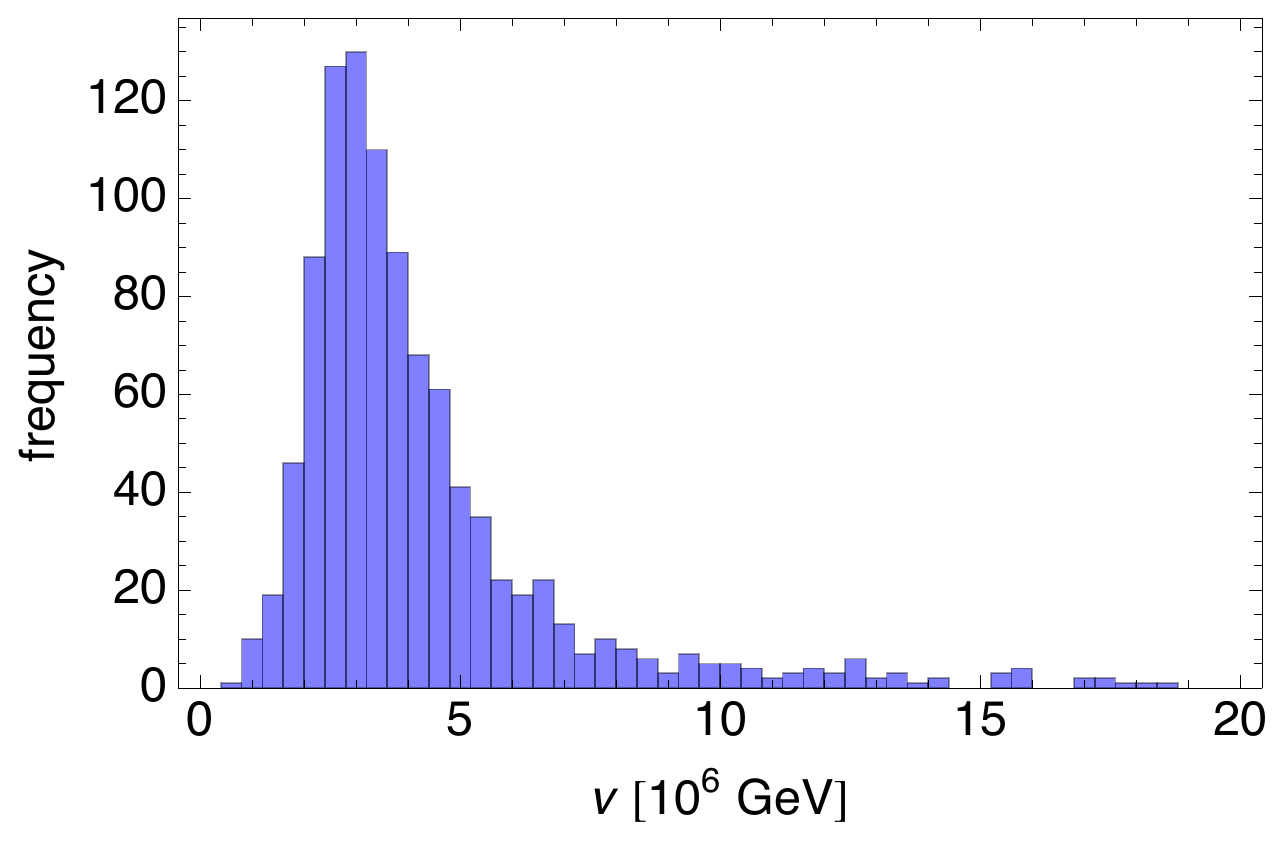}\quad \includegraphics[width=0.48\textwidth]{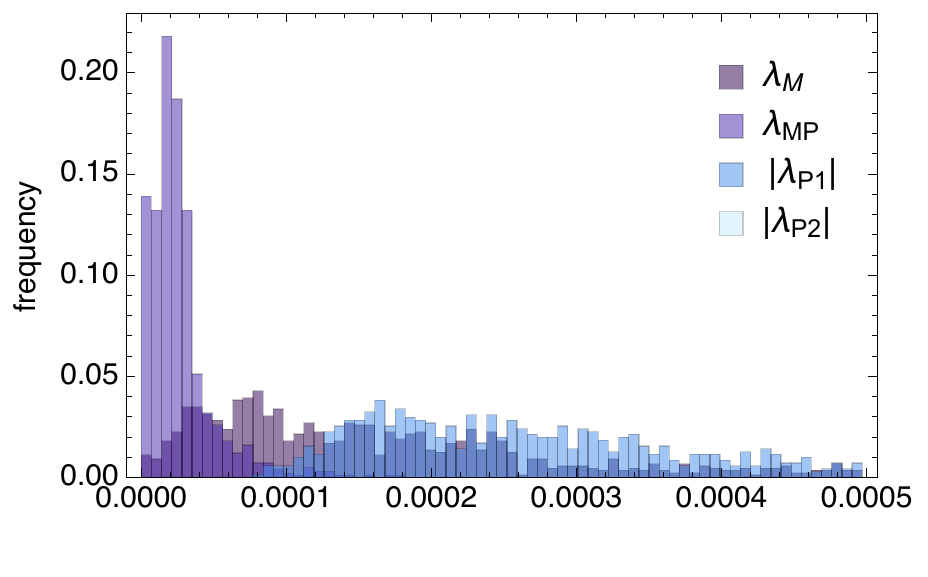}
	\end{center}
	\caption{{\bf Left panel:} Distribution of values of $v$ with $b=2.5\cdot 10^6$~GeV. 
	    {\bf Right panel:} Distribution of the quartic scalar couplings of 1000 viable 
	    scanned points with $b=2.5\cdot 10^6$~GeV.}
	\label{fig:vL}
    \end{figure}

In the numerical analysis we explore the parameter space by assuming an experimentally viable value for the leptocolour breaking 
scale, $b=2.5\cdot 10^6$~GeV. We demonstrate that a  small value of $\theta$ is preferred, and therefore the observed Higgs is mostly the pGB $\Pi_4$ with a tiny component of $\sigma$. The resulting values of the global symmetry breaking scale, $v$, and the distribution of quartic couplings are shown in Fig.\,\ref{fig:vL}.

We find that the preferred values of $v$ are roughly of the order of $b$, and this feature is reflected also in the Lagrangian mass parameters. To produce the correct Fermi scale this implies that tiny values of the angle $\theta$ are favoured.		
Furthermore, the values of the quartic couplings are overall very small, and in particular there is no large hierarchy between them.
This feature of tiny quartic couplings originates from the relation between the couplings of the scalar potential and the vacuum angle $\theta$ given by the minimisation conditions. In the limit of equal self-couplings, and $v=b$, this relation is given by $\lambda\sim\sin^2\theta$. Furthermore, the mass of the lightest scalar, i.e. the pGB Higgs, is in particular sensitive 
 to the coupling $\lambda_{MP}$, explaining the restricted values as shown in Fig.\,\ref{fig:vL}.
 
\section{Conclusions}

We have shown that the desired hierarchy between the EW and unification scales can be generated dynamically by radiative corrections in models with enhanced global symmetries and a pGB Higgs boson. We presented a minimal scenario utilising $\SU(4)\rightarrow \Sp(4)$ breaking pattern in the Pati--Salam framework. We find that a near alignment between the EW subgroup and the stability group is preferred resulting in global symmetry breaking scale near the leptocolour breaking scale. Furthermore, in this scenario all the scalar self-couplings are generally very small.

The presented scenario provides an interesting connection between the unification and the Fermi scale, and can be extended in various directions to incorporate e.g. dark matter, neutrino masses and mixings, or cosmic inflation.

\section*{Acknowledgments}

I thank the organizers of the 28th Rencontres de Blois for a fruitful conference. The CP$^3$-Origins centre is partially funded by the Danish National Research Foundation, grant number DNRF90. The author acknowledges partial funding from a Villum foundation grant.


\section*{References}

\begin{thebibliography}{99}

\bibitem{Georgi:1974sy}
  H.~Georgi and S.~L.~Glashow,
  Phys.\ Rev.\ Lett.\  {\bf 32} (1974) 438.
  doi:10.1103/PhysRevLett.32.438

\bibitem{Nishino:2012bnw}
  H.~Nishino {\it et al.} [Super-Kamiokande Collaboration],
  Phys.\ Rev.\ D {\bf 85} (2012) 112001
  doi:10.1103/PhysRevD.85.112001
  [arXiv:1203.4030 [hep-ex]].

\bibitem{Pati:1974yy}
  J.~C.~Pati and A.~Salam,
  Phys.\ Rev.\ D {\bf 10} (1974) 275
   Erratum: [Phys.\ Rev.\ D {\bf 11} (1975) 703].
  doi:10.1103/PhysRevD.10.275, 10.1103/PhysRevD.11.703.2

\bibitem{Parida:2014dba}
  M.~K.~Parida, R.~L.~Awasthi and P.~K.~Sahu,
  JHEP {\bf 1501} (2015) 045
  doi:10.1007/JHEP01(2015)045
  [arXiv:1401.1412 [hep-ph]].

\bibitem{Alanne:2014kea}
  T.~Alanne, H.~Gertov, F.~Sannino and K.~Tuominen,
  Phys.\ Rev.\ D {\bf 91} (2015) no.9,  095021
  doi:10.1103/PhysRevD.91.095021
  [arXiv:1411.6132 [hep-ph]].

\bibitem{Peskin:1980gc}
  M.~E.~Peskin,
  Nucl.\ Phys.\ B {\bf 175} (1980) 197.
  doi:10.1016/0550-3213(80)90051-6

\bibitem{Preskill:1980mz}
  J.~Preskill,
  Nucl.\ Phys.\ B {\bf 177} (1981) 21.
  doi:10.1016/0550-3213(81)90265-0

\bibitem{Galloway:2010bp}
  J.~Galloway, J.~A.~Evans, M.~A.~Luty and R.~A.~Tacchi,
  JHEP {\bf 1010} (2010) 086
  doi:10.1007/JHEP10(2010)086
  [arXiv:1001.1361 [hep-ph]].

\bibitem{Cacciapaglia:2014uja}
  G.~Cacciapaglia and F.~Sannino,
  JHEP {\bf 1404} (2014) 111
  doi:10.1007/JHEP04(2014)111
  [arXiv:1402.0233 [hep-ph]].

\bibitem{Alanne:2016mmn}
  T.~Alanne, H.~Gertov, A.~Meroni and F.~Sannino,
  arXiv:1608.07442 [hep-ph].

\bibitem{Gertov:2015xma}
  H.~Gertov, A.~Meroni, E.~Molinaro and F.~Sannino,
  Phys.\ Rev.\ D {\bf 92} (2015) no.9,  095003
  doi:10.1103/PhysRevD.92.095003
  [arXiv:1507.06666 [hep-ph]].

\bibitem{Alanne:2016uta}
  T.~Alanne, H.~Rzehak, F.~Sannino and A.~E.~Thomsen,
  arXiv:1606.03411 [hep-ph].

\bibitem{Alanne:2015fqh}
  T.~Alanne, A.~Meroni, F.~Sannino and K.~Tuominen,
  Phys.\ Rev.\ D {\bf 93} (2016) no.9,  091701
  doi:10.1103/PhysRevD.93.091701
  [arXiv:1511.01910 [hep-ph]].



\end{thebibliography}
\end{document}